\documentclass[aps,preprint]{revtex4}
\usepackage{amsfonts}
\usepackage{amsmath}
\usepackage{amssymb}
\usepackage{graphicx}

\input{tcilatex}

\begin{document}

\preprint{}
\title{\textbf{Modulation instability of lower hybrid waves leading to cusp
solitons in electron-positron-ion Thomas Fermi plasma}}
\author{\textbf{Zahida Ehsan }$^{1\ast }$, \textbf{N. L. Tsintsadze}$^{%
\mathbf{2}\ddag }$\textbf{\ and Renato Fedele}$^{\mathbf{3,4\dag }}$\textbf{%
\ }}
\affiliation{$^{1}$Department of Physics, COMSATS Institute of Information Technology,
Lahore 54000, Pakistan\\
$^{2}$Faculty of Exact and Natural Sciences and Andronicashvili Institute of
Physics, Javakhishvili Tbilisi University, Tbilisi 0128, Georgia.\\
$^{3}$Dipartimento di Fisica, Universit`a di Napoli \textquotedblright
Federico II\textquotedblright\ Complesso Universitario M.S. Angelo, Napoli,
Italy.\\
$^{4}$INFN Sezione di Napoli, Complesso Universitario di M.S. Angelo,
Napoli, Italy}

\begin{abstract}
Following the idea of three wave resonant interactions of lower hybrid waves
it is shown that quantum -modified lower hybrid (QLH) wave in electron
positron ion plasma with spatial dispersion can decay into another QLH wave
( where electron and positrons are activated whereas ions remain in the
background) and another ultra low frequency QULH (where ions are mobile).
Quantum effects like Bohm potential, exchange correlation and Fermi pressure
on the lower hybrid wave significantly reshaped the dispersion properties of
lower hybrid waves. Later a set of nonlinear Zakharov equations have been
derived to consider the formation of QLH wave solitons with the nonlinear
contribution coming from the QLH waves. Further, modulational instability of
the lower hybrid wave solitons is investigated and consequently it's growth
rates are examined for different limiting cases. Since the growth rate
associated with the three-wave resonant interaction are generally smaller
than the growth associated with the modulational instability, therefore only
latter have been investigated.

Soliton solutions from the set of coupled Zakharov and NLS equations in the
quasi-stationary regime have been studied. Ordinary solitons are attribute
of nonlinearity whereas a cusp soliton solution featured by nonlocal
nonlinearity have also studied.

Such an approach to lower hybrid waves and cusp solitons study in Fermi gas
comprising electron positron and ions is new and important. The general
results obtained in this quantum plasma theory will have widespread
applicability, particularly for processes in high energy plasma-laser
interactions set for laboratory astrophysics and solid state plasmas.

*For correspondence: ehsan.zahida@gmail.com

$\mathbf{\ddag }$nltsin@yahoo.com

\dag renato.fedele@na.infn.it
\end{abstract}

\author{}
\maketitle

\section{Introduction}

This paper is an attempt to develop a set of Zakharov equations for an
extremely high energy density matter, in particular a Fermi gas which makes
an assembly of electrons, positrons (holes) and ions. Here unlike usual pair
plasmas with same mass and charge \cite{1,2,3,4,5,6,7,8,9}, electrons and
holes (positrons) have mass asymmetry which is either due to interaction
between the particles or some other nonlinear phenomena emerging naturally
during the evolution of plasma could be at the root of this. On the other
hand small contamination of much heavier immobile ion, or small mass
difference of the electron positron (pair particles) can also produce
asymmetries in such plasmas. This mass asymmetry, however, opens up new and
interesting avenues for the plasma researchers because the physical
phenomena like waves and instabilities can now be scaled at both high and
low frequency times\cite{9}. 

Solid-state plasma or quantum (Fermi) liquid semiconductors are potential
applications where the effective mass of charge carriers (electrons and
holes) differs from that of free electrons. \ Charge carriers in
semiconductors like lighter electrons and heavier positive holes can make up
a degenerate system for example at $n_{e}\geq 10^{16}-10^{18}cm^{-3}$, with
the effective mass of electrons $m_{e}^{\ast }\approx (0.01-0.1)m_{e}$, and
at temperatures $T<10^{2}K$ \cite{10}. However mean free path is usually
longer than a few centimeters at low temperature in assembly of such highly
dense degenerate systems constituted by particles separated at a few
angstrom distances. And Pauli exclusion principle that permits collisions
only to final states, which were unoccupied before the collisions taking
place and screening of the Coulomb interaction between the particles are
strongly speculated for these long mean free paths\cite{11,12,13}. For this
reason, in our investigation we will suppose the plasma to be collisionless
at low temperatures. 

Quantum (degenerate) plasma characterize low temperature and high number
density system, can be found at various natural environments, for instance
in astrophysical \ environments (neutron stars, white dwarfs, magnatars
etc.), laser beam produced plasmas, nonlinear quantum optics,
microelectronic devices etc \cite{14}. 

Thermal de-Broglie wave length $(\lambda _{B}=\hbar /mv_{T})$ is a parameter
to determine the quantum degeneracy effects and entitles the spatial
extension of the wave function of constituent particles due to quantum
uncertainty and is either of the order or greater than the average
inter-fermionic distance, viz. ($d=n_{0}^{-1/3}$), where $n_{0}$ is the
equilibrium number density. In such scenario of high number densities, Fermi
pressure dominates over the thermal pressure which supports the compact
objects against the gravitational burst. 

While treating quantum plasma system, in the greater part of the existing
literature, Schrodinger, Pauli, Klein-Gordon, and Dirac like equation have
been cast into fluidized variables through Madelung transforms,
appropriately averaged to write down the fluid equations\cite{15,16}.

The idea is that the standard quantum equations of motion can be translated
into equivalent equations of \textquotedblleft classical\textquotedblright\
particles whose dynamics is determined by \textquotedblleft quantum
forces\textquotedblright\ (like the gradient of the Bohm potential) in
addition to the external forces\cite{15}.

Contrary to above Tsintsadze and Tsinsadaze\cite{17} developed a kinetic
equation for Fermi plasma using a single Fermi particle concept by employing
the non-relativistic Pauli equation and later with the aid of one-particle
distribution function. Where authors used one particle concept in spite of
the large number of particles in the unit volume all of them have only one
position and momentum $(r,p)$\cite{17}. And a huge literature exists on
linear investigations in Fermi plasmas\cite{18,19,20,21,22}. 

Furthermore, six years after the path-breaking paper published in 1972 by
Zakharov\cite{23}, Hidenori, Tomohiko, and Susumu in 1978 reported results
from the experiments on Modulation instability (MI) in electron plasma wave
which were precisely in agreement with the theory of Zakharov\cite{24}. They
found agreement between amplitude dependence of the modulation frequency in
the electron plasma wave and the modulation frequency dependence of the
amplitude. 

The modulational instability (MI) is a well-known mechanism for the energy
localization of wave packets in a nonlinear dispersive medium and can lead
to unstable situation which can potentially onset the formation stable
structures like envelope or cusp solitons or rogue waves in plasma like
medium. This happens when for example electromagnetic waves or light beam
decays eventually triggering the nonlinear structures. This kind of
instability has potential applications in nonlinear optics (lasers,
self-focusing, nonlinear radio waves, etc.), hydrodynamics, electromagnetic,
etc. that's why a huge number of work has been devoted to this\cite%
{25,26,27,28,29,30}. 

For instance, nonlinear properties of modulated 1-D drift-wave packets in a
inhomogeneous magnetized plasma were studied by Shukla et al \cite{27} using
nonlinear Schrodinger equation (NLSE), which depicts the formation of
drift-rogue waves and solitons (dark and bright ). Furthermore Liu et al.,
studied the nonlinear theory of cylindrical lower-hybrid drift-solitary
waves in an inhomogeneous, magnetized plasma using a two-fluid model and
reported attenuation in the wave amplitude and width of the solitary waves
with the increase in the inhomogeneity in density\cite{28}. Lower-hybrid
waves are well known to admit nonlinear structures, such as ordinary
solitons as well as envelope solitons which have been observed in the
Earth's magnetosphere by the FREJA satellite, and have been examined with
and without an extra charged species in plasma. Cusp type of soliton
structures have been paid less attention in plasmas; however, an attempt was
made by Ehsan, Tsintsadze et al., to study the decay of lower hybrid wave
into relatively less frequency lower hybrid wave in dusty plasmas eventually
forming the cusp type of solitons\cite{30}.

Since the study of these phenomena is also very important because of
possibility of heating and acceleration of plasmas by means of laser so
general results obtained in this quantum plasma theory will have widespread
applicability, particularly for processes in high energy plasma-laser
interactions. While in the environment of neutron stars pair plasmas are
speculated to be highly degenerate and ultradense that is why a rigorous
investigation for example in the frame work of quantum hydrodynamics\cite%
{31,32,33,34} of electron-positron ion degenerate plasma has been made over
the past few years.

In this study we investigate both the modulation instability of lower hybrid
wave, associated growth rates and stationary structures like cusp and
ordinary solitons in e-p-i Thomas Fermi plasma . 

This manuscript is organized in the following manner. In Sec. II, the basic
formulation of two types of lower-hybrid waves is given and the respective
dispersion relations are obtained. Sec. III deals with the mechanism of
three-wave resonant interaction, and the subsequent derivation of the
Zakharov and NLS equations is given in Sec. IV. In Sec. V, modulational
instabilities and associated growth rates are examined. Section VI
demonstrates the one dimensional analytical solutions of ordinary and cusp
solitons. Finally, main findings are recapitulated in Sec. VII. 

\section{Basic equations}

Considering the propagation of small longitudinal perturbations in an
electron-positron (hole)-ion plasmas, the relevant quantum Euler equations
for the $s$ species in quantum Fermi-Dirac plasmas \cite{35}:%
\begin{equation}
n_{s}\left( \frac{\partial }{\partial t}+\mathbf{v}_{s}\cdot \nabla \right) 
\mathbf{v}_{s}=\frac{q_{s}}{m_{s}}n_{s}\left( \mathbf{E}+\frac{\mathbf{v}_{s}%
}{c}\times \mathbf{B}_{o}\right) +\frac{\hbar ^{2}}{2m_{s}^{2}}\nabla \frac{1%
}{\sqrt{n_{s}}}(\nabla ^{2}\sqrt{n_{s}})-\frac{\mathbf{\nabla }P_{Fs}}{m_{s}}%
-\frac{\mathbf{\nabla }U_{s,xc}}{m_{s}},  \tag{1}
\end{equation}%
and 
\begin{equation}
\frac{\partial n_{s}}{\partial t}+\mathbf{\nabla \cdot (}n_{s}\mathbf{v}_{%
\mathbf{s}})=0.  \tag{2}
\end{equation}%
\begin{equation}
\Delta \varphi =4\pi e\left[ n_{e}-n_{p}-z_{i}n_{i}\right]   \tag{3}
\end{equation}
where propagation vector or wave vector ($\mathbf{k)}$ is taken along ${x}$%
-axis and the magnetic is in the ${z}$ direction $(B_{o}\mathbf{z}).$Last
term of Eq. (1) represents electron and positron exchange-correlation
potential which is a complex function of Fermi particles density and is
given as $U_{s,xc}=\frac{0.985e^{2}}{\epsilon }n_{s}^{1/3}\left[ 1+\frac{%
0.034}{a_{Bs}n_{s}^{1/3}}\ln \left( 1+18.37a_{Bs}n_{s}^{1/3}\right) \right] $%
\cite{36,37} is considered the attribute of the spin effects in dense
systems. For the readers it is useful to find that for the degenerate
plasma, these affects have been calculated comprehensibly in
\textquotedblleft Statistical Physics\textquotedblright\ book by Landau and
Lifshitz\cite{11} while exchange correlations for proton interaction have
been presented by Tsintsadze et al., \cite{38}. Since this depends upon the
number density, so we cannot ignore it in dense plasma environments. In Eq.
(1) $a_{Bs}=$\ $\epsilon \hbar ^{2}/m_{s}e^{2}$ is the well-known Bohr
atomic radius.

Equation (1) is general and conveniently written; however, later we will
treat ions as classical particle. In equation (1) $\hbar =h/2\pi $, $q_{j}$
the charge, $m_{j}$ mass and $c$ is the velocity of light in a vacuum of the 
$s$th species. Here, $s=i$ (ion), $s=e$ (electron), $s=p$ (positron or
holes), $q_{e}=-e$, $q_{p}=+e$ and $q_{i}=Z_{i}e$, with $e$ being the
magnitude of electronic charge and $Z_{i}$ is the number of charges on ions.
In Eq.(1), $P_{Fs}=\frac{\left( 3\pi ^{2}\right) ^{2/3}\hbar ^{2}n_{s}^{5/3}%
}{5m_{s}}$ is pressure law for 3-dimensional Fermi gas \cite{11}, and can
also be expressed in terms of Fermi energy such as $\frac{2}{5}\varepsilon
_{F}\left( n\right) n$ where $k_{B}$ is the Boltzmann constant, $T_{Fs}=%
\frac{\varepsilon _{F}}{K_{B}}=\frac{\hbar ^{2}\left( 3\pi ^{2}n_{0s}\right)
^{2/3}}{2m_{s}K_{B}}$\ is Fermi temperature, $n_{s}=n_{0s}+\delta n_{s}$ the
total number density with equilibrium number density $n_{0s}$\ and perturbed
number density $\delta n_{1s}$\ of $sth$ particles. \ The ion component can
be considered classical or quantum depending upon the relevant parameters.
However, in most of the situations, ions are considered as cold fluid while
describing the ion wave. In these dense quantum and semiclassical plasmas,
the screened interaction potential cannot be characterized by the standard
Debye-Huckel model according to the multiparticle correlations and the
quantum-mechanical effects such as the Bohm potential, quantum pressure, and
electron exchange terms since the average kinetic energy of the plasma
particle in quantum plasmas is of the order of the Fermi energy. Thermal
temperature of ions is small as compared to the electrons and positrons and
therefore ignored. 

\section{\textbf{Dispersion relation of quantum lower-hybrid wave }}

We linearize Eqs. (1-3) to obtain the linear and exact dispersion relation
of lower hybrid wave, and for this we use the solutions of plane waves when
we apply the plane wave solution we get the densities of electron and ions
which are perturbed and the quasineutrality condition for this case we
assume.

\begin{enumerate}
\item[\textbf{Case 1}] \textbf{Quantum modified lower hybrid wave (QLH)}
\end{enumerate}

In this case we assume ions stay in the background and that $\omega
_{ce}>\omega >\omega _{cp}$ and $n_{0i}-n_{i}$%
\begin{equation}
\delta n_{e}\simeq \delta n_{p}  \tag{4}
\end{equation}%
\begin{equation}
\frac{\delta n_{e}}{n_{oe}}=-\frac{\frac{e\varphi }{m_{e}}k_{0}^{2}}{\left[
\Omega _{ce}^{2}+k^{2}\left( V_{Fe}^{2}+\frac{\hbar ^{2}k^{2}}{4m_{e}^{2}}%
\right) \right] },  \tag{5}
\end{equation}%
and 
\begin{equation}
\frac{\delta n_{p}}{n_{op}}=\frac{\frac{e\varphi }{m_{p}}k_{0}^{2}}{\left[
\omega ^{2}-\Omega _{cp}^{2}-k^{2}\left( V_{Fp}^{2}+\frac{\hbar ^{2}k^{2}}{%
4m_{p}^{2}}\right) \right] }.  \tag{6}
\end{equation}%
\ where $\Omega _{cs}=\left( eB_{o}/m_{s}c\right) $ denotes the cyclotron
frequency of s species and $V_{Fs}^{2}=\frac{1}{3}\left( 3v_{Fs}^{2}-\alpha
_{s}-2\eta _{s}\right) $ represent combination of both Fermi velocity and
exchange correlation effect. There $v_{Fs}^{2}=\frac{6}{5}\frac{k_{B}T_{Fs}}{%
m_{s}}$ is the Fermi speed, $\alpha
_{s}=0.985(n_{s}^{1/3}e^{2}/m_{s}\varepsilon ),$ and $\eta
_{s}=1+(18.376n_{s}^{1/3}/a_{B})m_{s}\varepsilon .$

Now considering ions in the background and using the quasi-neutrality
condition $\delta n_{e}=\delta n_{p}$, we obtain a dispersion relation for
the lower hybrid waves propagating in electron positron ion plasma $\omega
^{2}\gg \omega _{cp}^{2}$ 
\begin{equation}
\omega ^{2}=\left[ \Omega _{lh}^{2}+\left( \frac{n_{0p}}{n_{0e}}\right)
\left( u_{Fs}^{2}+\frac{\hbar ^{2}k_{0}^{3}}{4m_{e}m_{p}}\right) k_{0}^{2}%
\right]   \tag{7}
\end{equation}%
where $\Omega _{lh}=\left[ \left( n_{po}/n_{eo}\right) \omega _{ce}\omega
_{cp}\right] ^{1/2}$ and $u_{Fs}=\left[ P_{Fs}^{2}/3m_{e}m_{p}\right] ^{1/2}$
lower-hybrid frequency and positron sound velocities in Fermi plasma
respectively. Whereas last term represents the Madelung contribution.
Deriving (7) we have assumed $m_{p}>m_{e}$ \ effective mass of electron is
less than the mass of positron. In semiconductors, for example, we often
have situations when the mass of the hole becomes much greater than the
effective mass of the electron \ and that $T_{Fp}<T_{Fe}$ and that $\omega
_{ce}>\omega >\omega _{cp}$ only the electrons are magnetized in the
derivation of expression (7). It is emphasized that (Eq. 7) has a spatial
dispersion term the contribution of which comes from the mass of positrons
and, as we shall see later, the spatial term can play an effective role in
the excitation of new modes. This propagation term also shows how the Debye
length is modified which shows that as the strength of magnetic field
increases, the Debye length $\left[ \lambda =2\pi v_{Fe}/\omega _{ce}\right] 
$ is the electron Thomas Fermi screening length. decreases, and thus the
particles in the Debye cloud remain mostly confined.

\begin{enumerate}
\item[Case 2] \textbf{Ultra low frequency lower hybrid wave (QULH) }
\end{enumerate}

For the dispersion relation this wave we assume ions are activated and the
quasi-neutrality condition reads as: 
\begin{equation}
\delta n_{e}=\delta n_{p}+\delta n_{i}  \tag{8}
\end{equation}%
or $\delta n_{p}+\delta n_{i}\simeq 0$ and under the condition $\omega
_{ce}\gg \omega \gg \omega _{cp}\gg $ we obtain

\begin{equation}
\omega _{L}^{2}=\left[ \Omega _{ulh}^{2}+\left( V_{si}^{2}+\frac{\hbar
^{2}k_{L}^{3}}{4m_{i}m_{p}}\right) k_{L}^{2}\right]   \tag{9}
\end{equation}%
where $\Omega _{ulh}=\sqrt{\left( n_{0i}/n_{0p}\right) \omega _{cp}\omega
_{ci}}$ and $V_{Fsi}^{2}=\left[ P_{Fi}^{2}/3m_{p}m_{i}\right] ^{1/2}.$In (9) 
$\omega _{L}\gg \omega _{ci}.$ For the readers it is interesting to note
that ions which are highly massive compared to electrons and positrons can
also be quantum if $T\leqslant T_{Fi}$ for example at $n_{i}\simeq
10^{22}cm^{-3},$ $m_{i}\simeq 10^{-24}g,$ $T_{Fi}\simeq 10^{0}Kelvin$ so
it's obvious that $T_{Fi}>T(\symbol{126}15$ or less).  

\section{\textbf{Excitation of QLH mode}}

Now using concept of nonlinear wave-wave interactions which are also known
as resonant wave-wave scattering or the decay instability, we consider the
possible decay of {the} quantum lower hybrid wave with frequency $\omega $
and wave vector $k$ into two waves, a QLH wave having frequency $\omega
^{^{\prime }}$ and wave number $k^{^{\prime }}$ and a QULH wave with
frequency $\omega _{L}$ and wave number $k_{L}$. This simple physical
picture can be obtained from Eqs.~(7) and (9), provided the energy and
momentum are conserved, i.e., 
\begin{align}
\omega -\omega ^{^{\prime }}& =\omega _{L}  \tag{10} \\
k-k^{^{\prime }}& =k_{L},  \notag
\end{align}%
where the components of momentum $k$, $k^{^{\prime }}$ and $k_{L}$ are
directed along the $x$-axis thus, $k$ and $k_{L}$ are scalars. From the
above relations, we obtain 
\begin{equation}
\omega -\omega ^{\prime }\simeq \left[ \left( \frac{n_{0i}}{n_{0p}}\right)
\omega _{cp}\omega _{ci}\right] ^{1/2}=\left[ \left( \frac{n_{p0}}{n_{e0}}%
\right) \frac{1}{\omega _{cp}\omega _{ce}}\right] ^{1/2}\left( v_{sp}^{2}+%
\frac{\hbar ^{2}k_{0}^{2}}{4m_{e}m_{p}}\right) k_{0}\left(
k_{0}-k_{0}^{\prime }\right) .  \tag{11}
\end{equation}%
We note here that we consider propagation in the $x$ direction only. Thus
using this simple model, we have shown {the} possibility of the three wave
interactions, which leads to the generation of the QULH waves. Here we do
not calculate growth rates for the three-wave resonant interaction but will
calculate more significant growth rates associated with the modulation
instability in the section after next.

\section{Construction of Zakharov equations}

Now for the excitation of the QULH mode, we will solve for the low frequency
density variations and include in our considerations the convective
derivative term $\left( \mathbf{\delta v}_{p}\cdot \nabla \mathbf{\delta v}%
_{p}\right) $, which leads to the ponderomotive force. We thus write down
the following equations for {the} positrons

\begin{equation}
\frac{\partial \mathbf{\delta v}_{p}^{ul}}{\partial t}+\left\langle \mathbf{%
\delta v}_{p}.\mathbf{\nabla \delta v}_{p}\right\rangle =\frac{e}{m_{p}}%
\left( \mathbf{E}^{ul}+\frac{1}{c}\mathbf{\delta v}_{p}^{ul}\times \mathbf{B}%
_{o}\right) -\frac{v_{Fp}^{2}}{n_{0p}}\frac{\partial \delta n_{p}^{ul}}{%
\partial x}+\frac{\hbar ^{2}}{4m_{p}^{2}}\frac{\partial ^{3}\delta n_{p}^{ul}%
}{\partial x^{3}}  \tag{12}
\end{equation}%
where the angular brackets denote the averaging over a typical lower-hybrid
wave period and wavelength, and $\mathbf{\delta v}_{p}$ is the positron
velocity for the QLH waves, $\mathbf{E}^{L}$ is the electric field for the
ultra-low frequency field. From the ion continuity equation we get%
\begin{equation}
\frac{\partial }{\partial t}\left( \frac{\delta n_{p}^{ul}}{n_{0p}}\right) +(%
\mathbf{\nabla }\cdot \delta \mathbf{v}_{p}^{ul})=0  \tag{13}
\end{equation}%
{The} ion dynamics are governed by the following equations of momentum and
continuity: 
\begin{equation}
\frac{\partial \mathbf{\delta v}_{\mathbf{i}}^{ul}}{\partial t}=-\frac{%
e\nabla \varphi ^{ul}}{m_{i}},  \tag{14}
\end{equation}%
and 
\begin{equation}
\frac{\partial }{\partial t}\left( \frac{\delta n_{i}^{ul}}{n_{0i}}\right) +(%
\mathbf{\nabla \cdot \delta v}_{\mathbf{i}}^{ul})=0.  \tag{15}
\end{equation}%
here we do not take into account ponderomotive force due to ions for being
relatively heavier than positrons will not be strong enough to cause
nonlinearity to appear in the ions dynamics. Using the quasi-neutrality
condition $\delta n_{p}^{ul}+\delta n_{i}^{ul}=0$, and after performing some
straightforward algebraic steps, we obtain a Zakharov-like equation
(Zakharov [1972])\cite{23,39}: 
\begin{equation}
\left( \frac{\partial ^{2}}{\partial t^{2}}+\Omega _{lh}^{2}-\frac{n_{0i}}{%
n_{0p}}v_{Fs}^{ul2}\frac{\partial ^{2}}{\partial x^{2}}+\frac{\hbar ^{2}}{%
4m_{p}m_{i}}\frac{n_{0i}}{n_{0p}}\frac{\partial ^{4}}{\partial x^{4}}\right) 
\frac{\delta n_{i}}{n_{0i}}=-\frac{m_{p}}{2m_{i}}v_{\phi }^{2}\frac{\partial
^{2}}{\partial x^{2}}\left\vert \frac{\delta n_{p}}{n_{0p}}\right\vert ^{2} 
\tag{16}
\end{equation}%
where $v_{\phi }=\left( \omega _{0}/k_{0}\right) .$ In the above equation
the that term is called source term which is due to the ponderomotive force
that comes from fast time scale in which positrons were involved and it
takes part to excite the QULH wave on slow time scale.

Now to obtain the second Zakharov's equation which is also famously known as
nonlinear Schrodinger equation, we start with the dispersion relation (7)
and treat $\omega $\ and $k$\ as operators given by $\left[ \omega =\omega
_{o}+i\frac{\partial }{\partial t}\right] $ and $\left[ k=k_{o}-i\frac{%
\partial }{\partial x}\right] $ \cite{30} and obtain%
\begin{equation}
i\left( \frac{\partial }{\partial t}+v_{g}\frac{\partial }{\partial x}%
-\alpha \frac{\partial ^{3}}{\partial x^{3}}\right) \delta n_{p}+\beta \frac{%
\mathbf{\partial }^{2}\delta n_{p}}{\partial x^{2}}-\Delta \omega \delta
n_{p}-\frac{\hbar ^{2}}{8m_{e}m_{p}}\frac{\partial ^{4}\delta n_{p}}{%
\partial x^{4}}+\omega _{_{\Gamma }}\frac{\delta n_{i}}{n_{0p}}\delta n_{p}=0
\tag{17}
\end{equation}%
where $n_{0p}^{o}$ and $n_{0e}^{o}$ are equilibrium densities of the
particles at fast time scale. $\Omega _{p0}=\left(
n_{0p}^{o}/n_{0e}^{o}\left( \omega _{ce}\omega _{cp}\right) \right) ^{1/2}$
and $\bigtriangleup \omega $ is the nonlinear frequency correction given by%
\begin{equation}
\bigtriangleup \omega =\frac{\frac{n_{0p}}{n_{0e}}\left[ \omega _{ce}\omega
_{cp}+k_{o}^{2}\left( u_{Fs}^{2}+\frac{\hbar ^{2}k_{o}^{4}}{4m_{p}m_{e}}%
\right) \right] -\omega _{o}^{2}}{2\omega _{o}};  \tag{18}
\end{equation}%
\begin{equation}
v_{g}=\frac{\partial \omega _{0}}{\partial k_{o}}=\left( \frac{n_{0p}}{n_{0e}%
}\frac{1}{\omega _{ce}\omega _{cp}}\right) ^{1/2}k_{0}\left( u_{Fs}^{2}+%
\frac{3k_{0}^{2}\hbar ^{2}}{2m_{p}m_{e}}\right) ;  \tag{19}
\end{equation}%
\begin{equation}
\alpha =\left( \frac{n_{0p}}{n_{0e}}\frac{1}{\omega _{ce}\omega _{cp}}%
\right) ^{1/2}\frac{\hbar ^{2}}{2m_{i}m_{p}}k_{0};  \tag{20}
\end{equation}%
and 
\begin{equation}
\beta =\frac{1}{2}\left( \frac{n_{0p}}{n_{0e}}\frac{1}{\omega _{ce}\omega
_{cp}}\right) ^{1/2}\left( u_{Fs}^{2}+\frac{3}{2}\frac{k_{0}^{2}\hbar ^{2}}{%
m_{p}m_{e}}\right)   \tag{21}
\end{equation}%
Equations~(16) and (17) together make set of Zakharov's like equation which
we will use in proceedings sections to study the modulation instability and
stationary structures.

\section{Modulational instability of QLH waves}

In this section, we investigate the modulational instability of the QLH
waves using Eqs.~(16 ) and (17) and to analyze the amplitude modulation of {%
the} QLH wave, we use Madelung's representation \cite{30} in the $x$%
direction only: 
\begin{equation}
\delta n_{p}\sim a(x,t)e^{iS(x,t)},  \tag{22}
\end{equation}%
where the amplitude $a$ and the phase $S$ are real, and substitution of (22)
into (16) and (17) gives real and imaginary parts, respectively. 
\begin{equation}
-a_{0}\left( \frac{\partial }{\partial t}+\mathbf{v}_{g}\cdot \frac{\partial 
}{\partial x}\right) \delta S+\beta \frac{\partial ^{2}}{\partial x^{2}}%
\delta a-\gamma \frac{\partial ^{4}}{\partial x^{4}}\delta a+\Omega
_{p0}a_{0}\frac{\delta n_{i}}{2n_{0p}}=0;  \tag{23}
\end{equation}%
\begin{equation}
\frac{\partial }{\partial t}\delta a+\mathbf{v}_{g}\frac{\partial \delta a}{%
\partial x}-\frac{\mathbf{1}}{2}\left( \frac{n_{0p}}{n_{0e}}\frac{1}{\omega
_{ce}\omega _{cp}}\right) ^{1/2}\frac{\hbar ^{2}}{m_{p}m_{e}}k_{0}\frac{%
\partial ^{3}\delta a}{\partial x^{3}}+\left( \frac{n_{0p}}{n_{0e}}\frac{1}{%
m_{ce}m_{cp}}\right) ^{1/2}\left( v_{Fs}^{2}+\frac{3\hbar ^{2}k_{0}^{2}}{%
2m_{p}m_{e}}\right) a_{0}\frac{\partial ^{2}\delta S}{\partial x^{2}}=0; 
\tag{24}
\end{equation}%
and%
\begin{equation}
\left( \frac{\partial ^{2}}{\partial t^{2}}+\Omega _{lh}^{2}-\frac{n_{0i}}{%
n_{0p}}u_{Fs}^{ul2}\frac{\partial ^{2}}{\partial x^{2}}+\frac{\hbar ^{2}}{%
4m_{p}m_{i}}\frac{n_{0i}}{n_{0p}}\frac{\partial ^{4}}{\partial x^{4}}\right) 
\frac{\delta n_{i}}{n_{0i}}=-\frac{m_{p}}{m_{i}}a_{0}\frac{\partial
^{2}\delta a}{\partial x^{2}}  \tag{25}
\end{equation}%
where $\gamma =\hbar ^{2}/8m_{p}m_{e}\sqrt{\omega _{ce}\omega _{cp}}.$To
this end, we linearize Eqs.(23-25) with respect to the perturbations, which
are represented as $a=a_{0}+\delta a$, $S=S_{0}+\delta S,$ where $a_{0\text{ 
}}$ and $S_{0}$ denote the equilibrium values whereas $\delta a$ and $\delta
S$ are small perturbations, $S_{o}=\Delta \omega t$.

We seek plane wave solution proportional to exp$\left[ i(\mathbf{k}_{L}\cdot 
\mathbf{r}-\omega _{L}t)\right] $ here $\mathbf{k}_{L}$ and $\omega _{L}$
are the wave number and frequency of the modulation. Finally, we obtain the
following dispersion relation for the modulation of quantum lower-hybrid
wave 
\begin{eqnarray}
&&\left[ \left( \omega _{L}-k_{L}\mathbf{v}_{g}\right) ^{2}-\alpha
k_{L}^{3}\left( \omega _{L}-k_{L}\mathbf{v}_{g}\right) -\beta
k_{L}^{2}\left( \beta k_{L}^{2}+\gamma k_{L}^{4}\right) \right] \left[
\omega _{L}^{2}-\Omega _{L}^{2}\right]   \notag \\
&=&\left( \frac{m_{e}}{m_{i}}\right) \left( \frac{n_{0p}n_{0i}}{n_{0e}^{2}}%
\right) \left( u_{Fs}^{2}+\frac{3}{2}\frac{\hbar ^{2}}{m_{p}m_{e}}\right)
k_{L}^{4}a_{0}^{2}  \TCItag{26}
\end{eqnarray}%
where $\Omega _{l}$ has been defined in the previous section. From (26), we
see that the diffraction term stabilizes the instability. For simplicity, we
discuss three limiting cases of the dispersion relation (26).

\subsection{Growth rates}

\textbf{Case 1: }First we will consider the case when $\omega _{L}\gg k_{L}%
\mathbf{v}_{g}$, $\alpha k_{L}^{3},$ $\beta k_{L}^{2}\left( \beta
k_{L}^{2}+\gamma k_{L}^{4}\right) $ and $\omega _{L}\gg \Omega _{L},$in this
case growth rate of  instability is 
\begin{equation}
\func{Im}\omega _{L}=\left( \frac{m_{i}}{m_{d}}\frac{n_{0p}n_{0i}}{n_{0e}^{2}%
}\right) ^{1/4}\left( v_{s}^{2}+\frac{3}{2}\frac{\hbar ^{2}}{m_{p}m_{e}}%
\right) ^{1/4}a_{0}^{1/2}k_{L}  \tag{27}
\end{equation}%
\textbf{Case 2: }Now we assume $\omega _{L}-k_{L}\mathbf{v}_{g}=\eta ,$ $%
\left( \omega _{L}^{2}-\Omega _{L}^{2}\right) \simeq \left( k_{L}^{2}\mathbf{%
v}_{g}^{2}-\Omega _{L}^{2}\right) .$ $\Omega _{L}^{2}>\left( k_{L}\mathbf{v}%
_{g}\right) ^{2}+\beta k_{L}^{2}\left( \beta k_{L}^{2}+\gamma
k_{L}^{4}\right) ,$ using this in (26), gives us%
\begin{equation}
\eta ^{2}=-\left( n_{0p}\omega _{ce}\omega _{cp}\right) ^{1/2}\frac{2\beta
m_{e}n_{0i}k_{L}^{4}a_{0}^{2}}{m_{i}n_{0e}^{3/2}\Omega _{L}^{2}}  \tag{28}
\end{equation}%
\textbf{Case 3: }We now consider a very important case that is at resonance $%
\omega _{L}-k_{L}\mathbf{v}_{g}=\eta ,\omega _{L}-\Omega _{L}=\eta ,$with
this (26) turns out to be 
\begin{equation}
\eta ^{3}-\alpha k_{L}^{3}\eta ^{2}-\beta k_{L}^{2}\left( \beta
k_{L}^{2}+\gamma k_{L}^{4}\right) \eta =\left( n_{0p}\omega _{ce}\omega
_{cp}\right) ^{1/2}\frac{2\beta m_{e}n_{0i}k_{L}^{4}a_{0}^{2}}{%
m_{i}n_{0e}^{3/2}}  \tag{29}
\end{equation}%
In this case growth rate is much larger. (29) is well-known cubic equation
of the form $ax^{3}+bx^{2}+cx+d=0$ which has three solutions.

\section{Soliton Solutions}

Stationary structures like solitons which are usually formed from
nonlinearly propagating waves have been rigorously investigated in plasmas
and other media\cite{10,26,40,41,42,43,44,45}. Here, we shall use the
standard approach to investigate solitons but we will restrict ourselves to
the consideration of stationary structures. Before we proceed further let us
assume in the second Zakharov's equation (17), $\beta \partial ^{2}/\partial
x^{2}(\delta n_{p})\gg \alpha \partial ^{3}/\partial x^{3}$ and $\hbar
^{2}/8m_{e}m_{p}\partial ^{4}/\partial x^{4}(\delta n_{p})$ and shifting to
a co-moving frame of reference $\xi =x-v_{g}t$ \ such that the perturbations
vanish at $\xi \rightarrow \pm \infty ,$ which yields%
\begin{equation}
\frac{\partial ^{2}\delta n_{p}}{\partial \xi ^{2}}-\frac{\Delta \omega }{%
\beta }\delta n_{p}-Q\delta n_{p}\frac{\partial ^{2}\delta n_{p}^{2}}{%
\partial \xi ^{2}}=0  \tag{30}
\end{equation}%
Using (30) and (16), we shall consider two types of soliton solutions.

\subsection{Ordinary solitons}

In the first case, we assume $\partial ^{2}/\partial t^{2}+\Omega
_{ulh}^{2}\ll \left( n_{0i}/n_{0p}\right) u_{Fs}^{ul2}\mathbf{\partial }%
^{2}/\partial x^{2}$ and from the Zakharov's equations (16), we obtain the
following expression for the perturbed density:

\begin{equation}
\delta n_{i}=\frac{m_{p}}{2m_{i}}\frac{v_{\varphi }^{2}}{v_{s}^{ul2}}\frac{%
\left\langle \delta n_{p}^{2}\right\rangle }{n_{0p}}  \tag{31}
\end{equation}%
Using (13) and (30) and introducing the notations $P=\left( \frac{%
Z_{d}^{2}n_{od}}{m_{d}}\right) \left( \frac{m_{i}}{m_{d}}\right) \omega
_{_{\Gamma }}\delta v_{i}$, and $Y=2/v_{o}\left( \omega _{_{\Gamma
}}\triangle \omega \right) ^{1/2}\xi $, and then integrating once, we get 
\begin{equation}
\frac{dP}{dX}=P\sqrt{1-P^{2}}  \tag{32}
\end{equation}%
which has a solution of the form 
\begin{equation}
Y=\log \left[ \frac{C}{1+\sqrt{1-C^{2}}}\right] .  \tag{33}
\end{equation}%
This is a standard bright soliton structure. It is important to mention, we
have neglected higher order terms which could be interesting but for the
solution we need to compute the equations which is beyond the scope of
present investigation. 

\subsection{Cusp soliton}

Here we assume that first and third terms of (16) compensate each other and
so the following expression for the perturbed ion density comes out 
\begin{equation}
\frac{\delta n_{i}}{n_{0i}}\approx -\frac{1}{2k_{0}^{2}}\frac{\partial ^{2}}{%
\partial x^{2}}\left( \frac{\delta n_{p}^{2}}{n_{0i}n_{0e}}\right)  \tag{34}
\end{equation}%
Here we notice {the} interesting feature that the density perturbation is
proportional to the second derivative of the amplitude of {the} density
which in turn leads to a soliton of cusp kind. Substituting relation (34) in
the second Zakharov equation (31), we obtain 
\begin{equation}
\left( \frac{dC}{dz}\right) ^{2}\left( 1-2Q^{\prime }C^{2}\right) =C^{2} 
\tag{35}
\end{equation}%
where $C=\delta n_{e}/n_{0p}$, $z=\xi \sqrt{\Delta \omega /\beta }$ and%
\begin{equation}
Q^{\prime }=\frac{\omega _{ce}\omega _{cp}}{k_{0}^{2}\left(
v_{s}^{2}+3k_{0}^{2}\hbar ^{2}/2m_{_{p}}m_{e}\right) }\left(
n_{0p}/n_{0e}\right)  \tag{36}
\end{equation}%
and further by introducing $\Psi =\sqrt{2Q^{\prime }}C$ and changing the
variable from $\xi $ to $Y$ and integrating Eq.~(35) once, we obtain equation

\begin{equation}
\frac{d\Psi }{dY}\sqrt{1-\Psi ^{2}}=\Psi   \tag{37}
\end{equation}%
above equation yields that when $\Psi $ reaches at maximum value $(\Psi =1)$%
, the first derivative goes to infinity $(d\Psi /dY\rightarrow \infty )$,
which shows the evidence of cusp solitons formation. The second integration
of Eq.(37) leads to the following solution 
\begin{equation}
Y=\sqrt{1-\Psi ^{2}}+\log \mid \Psi \mid -\log \mid 1+\sqrt{1-\Psi ^{2}}\mid 
\tag{38}
\end{equation}%
The graph [Figure] of which exhibits an infinite discontinues slope (cusp)
at its crest, and which justifies the term \textquotedblleft cusp
soliton\textquotedblright , but all physical quantities remain continues.
This type of soliton is quite different from the well known smooth solitons
because of having infinite1st derivative at its maxima. 

\section{Conclusions}

During this research a theoretical model {was} presented for the excitation
of ultra-low frequency quantum-lower hybrid mode (QULH) oscillating at
frequency. {This model employs} the decay of a relatively high frequency
modified lower-hybrid (QLH) wave into a relatively lower frequency QLH and
QMLH based on {three}-wave resonant interaction. Modulational instabilities
of QLH waves are investigated and {their} growth rates are studied.
Additionally, one-dimensional nonlinear localized structures of bright
solitons and nonlinear nonlocal structures like cusp solitons are obtained. {%
The} generation of cusped solitary waves is considered by the modulation of
the lower hybrid wave amplitude. We feel that such an approach to lower
hybrid waves is important and demands attention.


\begin{thebibliography}{99}
\bibitem{1} F. Pacini, Nature London 219, 145 (1968).

\bibitem{2} P.  Goldreich and W. H. Julian, Astrophys. J. 157, 869 (1969).

\bibitem{3} M. J. Rees, Nature London 229, 312 (1971).

\bibitem{4} C. M. Surko, M. Leventhal, and A. Passner, Phys. Rev. Lett. 62,
901 (1989).

\bibitem{5} H. Boehmer, M. Adams, and N. Rynn, Phys. Plasmas 2, 4369 (1995).

\bibitem{6} V. I. Berezhiani, S. M. Mahajan, and N. L. Shatashvili, Phys.
Rev. A 81, 053812 (2010).

\bibitem{7} V. I. Berezhiani, S. M. Mahajan, and N. L. Shatashvili, J.
Plasma Phys. 76, 467 (2010).

\bibitem{8} P. Liang, S. C. Wilks, and M. Tabak, Phys. Rev. Lett. 81, 4887
(1998);S.A. Khan, M. Ilyas, Z. Wazir and Z. Ehsan, Astrophys Space Sci DOI
10.1007/s10509-014-1925-8.

\bibitem{9} Z. Ehsan, N. L. Tsintsadze, H. A. Shah, R. M. G. M. Trines, and
M. Imran Phys. Plasmas 23, 062125 (2016) and refernces therein.

\bibitem{10} N. L.Tsintsadze, L.N. Tsintsadze, A. Hussain, and G. Murtaza
Eur. Phys. J. D 64, 447--452 (2011).

\bibitem{11} L. D. Landau, E.M. Lifshitz, Statistical Physics, 3rd edn.,Part
1 (Butterworth-Heinemann, Oxford, 1998).

\bibitem{12} D. Kremp, M. Schlanges, W.D. Kraeft, Quantum Statistics of
Nonideal Plasmas (Springer-Verlag, Berlin Heidelberg,2005).

\bibitem{13} N. L. Tsintsadze, A. Rasheed, H.A. Shah, G. Murtaza, Phys.
Plasmas 16, 112307 (2009).

\bibitem{14} P. K. Shukla, B. Eliasson, Phys. Usp. 53, 51 (2010), and
references therein.

\bibitem{15} S. M. Mahajan and F. A. Asenjo Phys. Plasmas 23, 056301 (2016) 

\bibitem{16} G. Manfredi, Fields Inst. Commun. 46, 263 (2005); F. Haas,
Europhys. Lett. 77, 45004 (2007).

\bibitem{17} N. L. Tsintsadze and L. N. Tsintsadze, Europhys. Lett., 88,
35001 (2009); N.L. Tsintsadze, L.N. Tsintsadze, in From Leonardo to ITER:
Nonlinear and Coherence Aspects, AIP Proc. No. CP1177 edited by J. Weiland
(AIP, New York, 2009) 18.

\bibitem{18} I. I. Goldman, Zh. Eksp. Teor. Fiz. 17, 681 (1947).

\bibitem{19} Y. L. Klimontovich, V.P. Silin, Zh. Eksp. Teor. Fiz. 23, 151
(1952).

\bibitem{20} D. Bohm, Phys. Rev. 85, 166 (1952)

\bibitem{21} D. Bohm, D. Pines, Phys. Rev. 92, 609 (1953).

\bibitem{22} D. Bohm, D. Pines, in Plasma Physics, edited by J.E. Drummond
(McGraw-Hill, New York, 1961).

\bibitem{23} V. E. Zakharov, Zh. Eksp. Teor. Fiz. 62, 1745 (1972).

\bibitem{24} A. Hidenori, Y. Tomohiko, and T. Susumu, Electric. Eng. Jpn.
98, 2(1978).

\bibitem{25} N. L. Tsintsadze, Z. Ehsan, H. A. Shah and G. Murtaza, Phys.
Plasmas 13,072103 (2006).

\bibitem{26} Z. Ehsan, N. L.Tsintsadze, J. Vranjes, and S. Poedts, Phys.
Plasmas16, 053702 (2009).

\bibitem{27} P. K. Shukla and R. K. Varma, Phys. Fluids B 5, 236 (1993).

\bibitem{28} H-F Liu, S-Q Wang, K-H Li, Z-H Wang, W-B Zhang, Z-L Wang,
Q-Xiang, K-Huang, Y-Liu, S-Li, F-Z Yang and L-Chang, Phys. Plasmas 20,
044502 (2013).{}

\bibitem{29} A. P. Misra, Physics of Plasmas 21, 042306 (2014). 

\bibitem{30} Z. Ehsan, N. L. Tsintsadze, H. A. Shah and G. Murtaza Physics
of Plasmas 16, 023702 (2009).

\bibitem{31} F. Haas, L.G. Garcia, J. Goedert, Phys. Plasmas 10, 3858
(2003); F. Haas, J. Plasma Phys. 79, 371 (2013).

\bibitem{32} A. Mushtaq, S.A. Khan, Phys. Plasmas 14, 052307 (2007); F.
Haas, B. Eliasson, and P. K. Shukla, Phys. Rev. E 86, 036406 (2012).

\bibitem{33} E. F. El-Shamy, R. Sabry, W.M. Moslem, P.K. Shukla, Phys.
Plasmas 17, 082311 (2010); B. Eliasson and P. K. Shukla, Phys. Rev. E 83,
046407 (2011).

\bibitem{34} A. S. Bains, A.P. Misra, N.S. Saini, T.S. Gill, Phys.Plasmas
17, 012103 (2010).

\bibitem{35} Z. Ehsan, M. Shahid, M. A. Rana, M. Ahmad, A. Abdikian and A.
Shahbaz arXiv:1709.08123 [physics.plasm-ph].

\bibitem{36} A. Abdikian and Z. Ehsan Phys Lett A 3 81 2939 (2017) and
refernces therein.

\bibitem{37} N. Crouseilles, P. A. Hervieux, and G. Manfredi, Phys. Rev. B
78, 155412 (2008).

\bibitem{38} N. L. Tsintsadze, G. Murtaza and Z. Ehsan, Phys. Plasmas 13,
22103 (2006).

\bibitem{39} R. Fedele, P. K. Shukla, M. Onorato, D. Anderson, and M. Lisak,
Phys. Lett. A 303, 61 (2002).

\bibitem{40} V. I. Karpman, NonlinearWaves in Dispersive Media (Pergamon
Press, Oxford, 1975). 

\bibitem{41} R. Z. Sagdeev, In: Leontovich, M.A. (ed.) Reviews of Plasma
Physics (Consultants Bureau, New York,1966), vol. 4, p. 23.

\bibitem{42} C. Sulem and P. L. Sulem, The Nonlinear Schr%
\"{}%
odinger Equation: Self-focusing and Wave Collapse (Springer, New York, 1999).

\bibitem{43} A. Mannan, R. Fedele, M. Onorato, S. De Nicola and D. Jovanovi
Phys Rev E 91, 012921 (2015).

\bibitem{44} O. R. Rufai, A.S. Bains, and Z. Ehsan Astrophys Space Sci
357:102 (2015) and refernces therein.

\bibitem{45} A. Sohail, S. Arshad and Z. Ehsan, Int. J. Appl. Comput. Math
(2017). https://doi.org/10.1007/s40819-017-0420-7 and refernces therein.
\end{thebibliography}
\end{document}